# A transportable hyperspectral imaging setup based on fast, high-density spectral scanning for *in situ* quantitative biochemical mapping of fresh tissue biopsies

**Luca Giannoni,[a,b,†,*] Marta Marradi,[a,b,†] Kevin Scibilia,[c] Ivan Ezhov,[c] Camilla Bonaudo,[d] Angelos Artemiou,[e] Anam Toaha,[a,b], Frédéric Lange,[e] Charly Caredda,[f] Bruno Montcel,[f] Alessandro Della Puppa,[d] Ilias Tachtsidis,[e] Daniel Rückert,[c,g] and Francesco Saverio Pavone[a,b,h]**

[a] University of Florence, Department of Physics and Astronomy, Florence, Italy
[b] European Laboratory for Non-Linear Spectroscopy, Sesto Fiorentino, Italy
[c] Technical University of Munich, TranslaTUM - Center for Translational Cancer Research, Munich, Germany
[d] Azienda Ospedaliero-Universitaria Careggi, University of Florence, Neurosurgery, Department of Neuroscience, Psychology, Pharmacology and Child Health, Florence, Italy
[e] University College London, Department of Medical Physics and Biomedical Engineering, London, UK
[f] Univ Lyon, INSA-Lyon, Université Claude Bernard Lyon 1, UJM-Saint Etienne, CNRS, Inserm, CREATIS UMR 5220, Lyon, France
[g] Imperial College London, Department of Computing, London, UK
[h] National Research Council, National Institute of Optics, Sesto Fiorentino, Italy

**Abstract**

**Significance:** Histopathological examination of surgical biopsies, such as in glioma and glioblastoma resection, is hindered in current clinical practice by the long times required for the laboratory analysis and pathological screening, typically taking several days or even weeks to be completed.

**Aim:** We propose here a transportable, high-density, spectral-scanning based hyperspectral imaging (HSI) setup, named HyperProbe1, that can provide *in situ*, fast biochemical analysis and mapping of fresh surgical tissue samples, right after excision, and without the need of fixing, staining nor compromising the integrity of the tissue properties.

**Approach:** HyperProbe1 is based on spectral scanning via supercontinuum laser illumination filtered with acousto-optic tuneable filters. Such methodology allows the user to select any number and type of wavelength bands in the visible and near-infrared range between 510 and 900 nm (up to a maximum of 79), and to reconstruct 3D hypercubes composed of high-resolution (4-5 μm), widefield images (0.9x0.9 $mm^2$) of the surgical samples, where each pixel is associated with a complete spectrum.

**Results:** The HyperProbe1 setup is here presented and characterised. The system is applied on 11 fresh surgical biopsies of glioma from routine patients, including different grades of tumour classification. Quantitative analysis of the composition of the tissue is performed via fast spectral unmixing to reconstruct mapping of major biomarkers, such as oxy- ($HbO_2$) and deoxyhaemoglobin (HHb), as well as cytochrome-c-oxidase (CCO). We also provided a preliminary attempt to infer tumour classification based on differences of composition in the samples, suggesting the possibility to use lipid content and differential CCO concentrations to distinguish between lower and higher grade gliomas.

**Conclusions:** A proof-of-concept of the performances of HyperProbe1 for quantitative, biochemical mapping of surgical biopsies is demonstrated, paving the way for improving current post-surgical, histopathological practice via non-destructive, *in situ* streamlined screening of fresh tissue samples in a matter of minutes after excision.

* Luca Giannoni (corresponding author), E-mail: giannoni@lens.unifi.it
† The authors contributed equally

## 1 Introduction

Histopathological screening of excised tissue is the current "gold standard" in post-surgical oncological practice[1], for clinical and molecular evaluation of critical parameters such as type, grading and classification of tumours, e.g., in glioma and glioblastoma (GBM) resection [2–4]. Normal routine involves the dispatch of fresh surgical biopsies after resection to the histopathology laboratory: there, the samples are typically fixed for preservation, sectioned, stained -various staining techniques are used, with haematoxylin and eosin (H&E) staining being the most prominent- and then imaged with a microscope to determine their structural and molecular composition[3]. However, modern histopathological analysis presents several limitations, the most severe one being the lengthy preparation of the samples that leads to long duration of the procedures to obtain the final results, which can vary from several days to even weeks after the surgery. Extemporaneous and intraoperative analyses can be much faster (minutes to hours), but the number of biopsy samples is limited due to operational logistics and costs of the procedures, and the breadth of information they can provide is very limited for diagnostic purposes[5]. Furthermore, diagnosis and classification can be affected by variability in the subjective interpretation of the results by histopathologists, with the screening essentially lacking a more quantitative and objective way to systematically process the imaging outcomes[6]. Overall, post-operative prognosis and planning would enormously benefit from a different approach to histopathology that could provide much faster and more reliable information on the tissue biopsies, ideally by having a screening *in situ* right after the surgery that could lead to quantitative results in a matter of minutes to hours.

Hyperspectral imaging (HSI) is an optical imaging modality that is becoming increasingly more notable in recent years in the biomedical and bioimaging fields[7], and whose main features can be particularly suited and advantageous to tackle the abovementioned challenge. HSI acquires and reconstructs images of a target at multiple, narrow, contiguous or adjacent wavelength bands in the electromagnetic spectrum, typically spanning from the visible to the near infrared (NIR) range[8]. This allows the user to obtain 3D spatio-spectral datasets, named “hypercubes”, where each spatial pixel of the images is associated with a corresponding spectrum of reflected, transmitted and/or fluorescent light. The information carried by the hypercubes is related to the optical properties of absorption and scattering of the investigated tissue, from which is then possible to infer, map and quantify its biochemical and structural composition, without the need for time-consuming staining procedures or the use of any exogenous contrast agent. Intrinsic biomarkers for physiology and pathophysiology of the tissue can indeed be identified for diagnostic purposes, such as haemoglobin for haemodynamics, oxygenation and vascularisation, or cytochrome-c-oxidase (CCO) for cellular metabolism[8–10], and related to tumour key parameters of classification[11]. In addition to its capabilities for non-destructive biochemical analysis of freshly excised tissue, HSI has the additional advantage of fast image acquisition and data processing, mainly thanks to recent advancements in deep learning and artificial intelligence (AI) algorithms[12], achieving near real-time computing and almost immediate visualization of the results[13]. Finally, HSI technology is typically compact and relatively inexpensive (compared to other traditional imaging modalities), so that devices can be developed to be fully transportable and capable to easily fit either within the surgical room or in its proximity (e.g., in a post-surgical area) without encumbrance.

We present here the first prototype of a compact, fully transportable HSI setup called “HyperProbe1”, which is able to rapidly select at high density, sampling and spectral resolution (3.5 to 7 nm of minimum bandwidth) any desired wavelength band between 510 and 900 nm, and to image a target at a field of view (FOV) of 0.9x0.9 $mm^2$ with up to 79 spectral bands, in less than 5 minutes. HyperProbe1 has the capability to image broadband reflected light from fresh biopsies and to reconstruct maps of their optical properties, as well as to quantify the content of biomarkers of interest within the examined tissue (such as the two forms of haemoglobin and CCO) via fast spectral unmixing algorithms. We provide full technical characterisation of the performances of HyperProbe1 and a proof-of-concept of its application on samples of freshly excised glioma from surgical biopsies at different World Health Organization (WHO) gradings[14]. The success of HyperProbe1 in providing quantitative biochemical analysis and mapping of surgical biopsies can pave the way to a novel, fast and heightened methodology to perform *in situ* histopathological screening right after surgery, without the need for any manipulation or degradation of the samples.

## 2 Material and methods

### *2.1 The HyperProbe1 system*

HyperProbe1 is a HSI system based on spectral scanning acquisition mode, where the target is illuminated in rapid sequence at each selected wavelength band, whilst a full-frame image is acquired synchronously at each illumination step.

**Table 1** List of components of HyperProbe1 and their specifications.

| Component | Manufacturer & model | Key specifics |
|---|---|---|
| SCL | NKT Photonics,<br>SuperK FIANIUM FIR20 | Broadband illumination (400-2400 nm);<br>Maximum total power of 6.5 W; |
| AOTF | NKT Photonics,<br>SELECT VIS-nIR | Broadband selection (510-900 nm);<br>Spectral resolution of 3.5-7 nm (FWHM); |
| Camera | Hamamatsu,<br>ORCA-Flash 3.0 | 4.2-MP (2048 x 2048) CMOS sensor;<br>6.5-µm pixel size;<br>QE up to 82% (Visible and NIR);<br>Maximum frame rate of 40 fps; |
| Amplitude stabilisers | Thorlabs;<br>NEL02A/M + NEL03A/M | Amplitude stabilisation within ±0.05%; |
| Speckle reducer | Optotune; LSR-3005-6D-NIR | Transmission up to 98%; |
| Optic fibres | NKT Photonics,<br>SuperK CONNECT | Broadband coverage (400-2000 nm);<br>High power throughput (up to 500 mW);<br>1-mm core diameter; |
| Objective | Thorlabs,<br>LMM15X-P01 | 15x reflective objective;<br>NA = 0.3; |

CMOS: Complementary metal-oxide semiconductor; FWHM = Full-width at half maximum; QE = Quantum efficiency; NIR = Near-infrared; FOV = Field of view; NA = Numerical aperture.

All the spectral frames are then stacked together to reconstruct the corresponding 3D hypercube of the target[8]. A schematics of the configuration of HyperProbe is reported in Fig.1a, whereas a detailed list of its components is presented in Table 1.

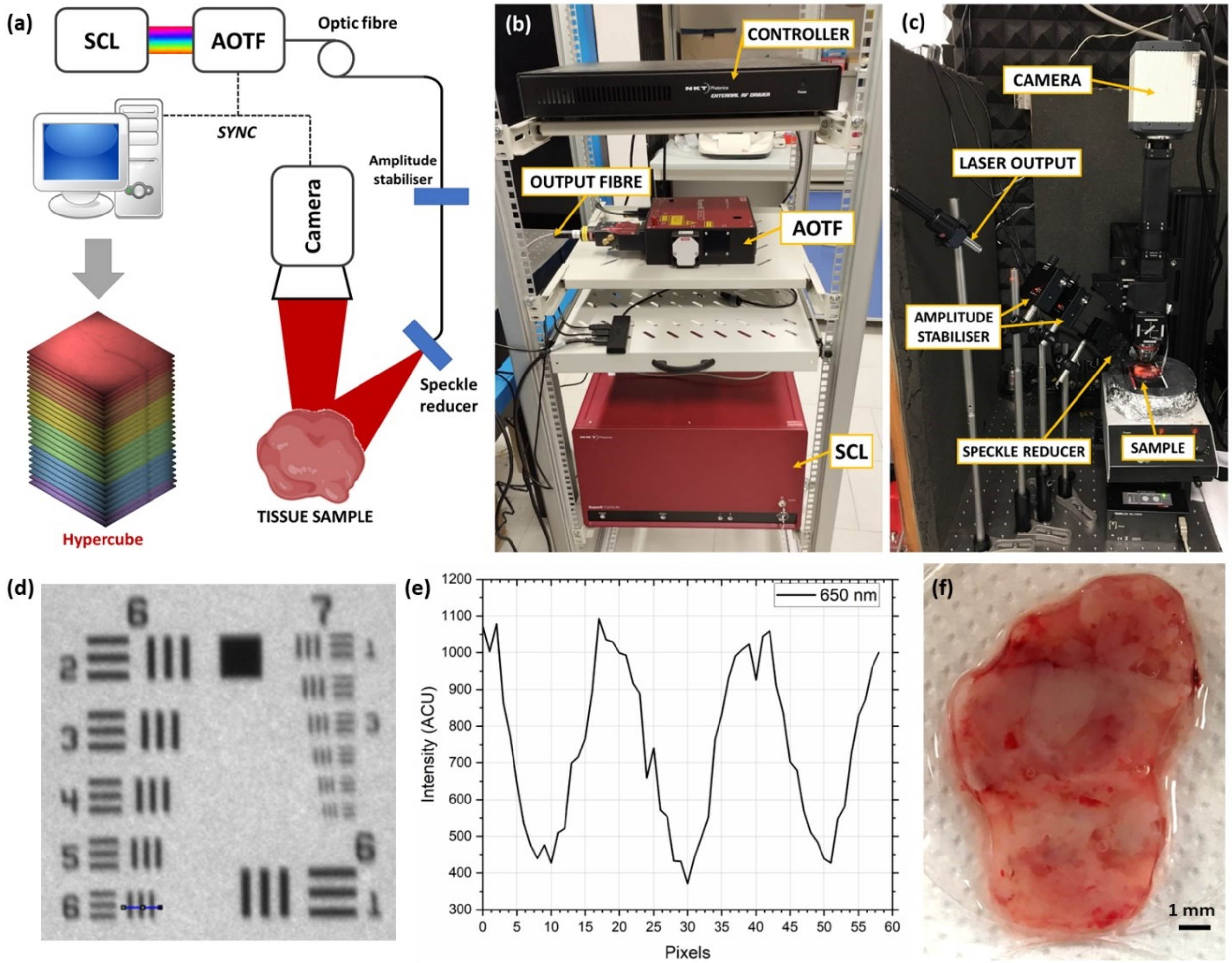

**Fig. 1** (a) Schematics of the HyperProbe1 with all its components; (b) Picture of the spectral illumination side of HyperProbe1, including the SCL source, the AOTF and the controlling devices; (c) Picture of the imaging and detection side of HyperProbe1, depicting the illumination output, the amplitude stabilisers, the speckle reducer and the camera; (d) Result of the imaging tests on the USAF1951 target with HyperProbe1 at 650 nm; (e) The smallest resolved line pair (group 6, element 6) highlighted in blue and its corresponding line profile that shows minimum FWHM separation (4.38 μm); (f) Example of a sample of excised glioma tissue obtained from the surgical biopsies.

The illumination side of HyperProbe1 (Fig. 1b) is composed of a supercontinuum laser[15] (SCL;

NKT Photonics, SuperK FIANIUM FIR20), generating a coherent, broadband illumination (400-2400 nm) at a maximum total power of 6.5 W, and a set of acousto-optic tuneable filters (AOTF; NKT Photonics, SELECT VIS-nIR) that selectively filters any desired spectral band between 510 and 900 nm, with a minimum bandwidth (full width at half maximum; FWHM) of 3.5 nm in the visible and a maximum bandwidth of 7 nm in the NIR range. The filtered light from the AOTF is directed to the sample through an optic fibre delivery system (NKT Photonics, SuperK CONNECT) of 1-mm core diameter, coupled with: (1) a pair of laser amplitude stabiliser connected in series (Thorlabs, NEL02A/M and NEL03A/M), to eliminate intensity noise and maintain illumination stability over time within 0.05% of a selected output power across the whole spectral range, (2) a laser speckle reducer (Optotune, LSR-3005-6D-NIR), and (3) an achromatic doublet lens, to make the beam divergent in order to obtain an illumination spot of 2-3 $cm^2$ on the target. Image acquisition at each spectral band is obtained on the imaging side of HyperProbe1 (Fig. 1c) by use of a complementary metal-oxide semiconductor (CMOS) camera (Hamamatsu, ORCA-Flash 3.0). The CMOS camera has a sensor size of 2048 x 2048 pixels, with pixel size of 6.5 μm, 82% peak quantum efficiency (at 620 nm), and maximum readout rate of 40 fps. It is coupled with a 15x reflective objective (Thorlabs, LMM-15X-P01) and an infinity-corrected, conjugated tube lens (Thorlabs, TTL200-B), to generate a FOV of the target of about 0.9 x 0.9 $mm^2$. The whole setup is mounted on a wheeled rack (as depicted in Fig.1b) and can be easily transported within or adjacent to the surgical room, with the illumination probe and imaging side that can be laid on a small breadboard for improved stability.

The HyperProbe1 system was characterised in terms of power emission, spectral, temporal and imaging performances. The complete characteristics and features of HyperProbe1 are reported in Table 2. HyperProbe1 can scan the entire spectral range of operation (510-900 nm) by sampling sequentially up to 79 wavelength bands at 5-nm steps. Each spectral band is also modulated in amplitude by the AOTF, to provide an approximately constant output power on the target of about 200-450 μW per band, accounting also for the quantum efficiency (QE) of the camera. This is aimed at maintaining a fixed integration time of the camera for each spectral frame (between 5 and 30 ms, depending on the target), as well as an adequate signal-to-noise ratio (SNR) throughout every acquired spectral frame. The typical acquisition time for an entire hypercube (79 bands) ranges between 1 to 5 minutes (depending on the selected integration time of the camera), for biological targets. Spatial resolution of HyperProbe1 was assessed for each single spectral band via imaging of a positive resolution test target, the USAF 1951 (Thorlabs, R1L1S1P). For all these imaging assessments, the spatial resolution of HyperProbe1 was found equal to 114 lp/mm, corresponding to a smallest resolvable detail of 4.38 μm (as shown in Fig. 1d and Fig. 1e, for 650 nm).

**Table 2** Technical characteristics and features of HyperProbe1.

| **Characteristics** | **Illumination side** |
|---|---|
| Illumination mode: | Spectral scanning |
| Available spectral range: | 510-900 nm |
| Minimum sampling step size: | 5 nm |
| Maximum number of spectral bands: | 79 (visible and NIR) |
| Spectral resolution (FWHM): | 3.5 nm (visible),<br>7 nm (NIR) |
| Average output power per spectral band: | ~200 μW (visible),<br>~450 μW (NIR) |
| Power stability over time: | ± 0.05% |
| **Characteristics** | **Imaging side** |
| Type of detector: | CMOS |
| Sensor format: | 2048 x 2048 pixels |
| Pixel size: | 6.5 μm |
| Spatial resolution: | 4.38 μm |
| FOV: | 0.9 x 0.9 $mm^2$ |
| Frame rate: | 40 fps (at full format) |
| Sensitivity (QE): | 82% at 620 nm |
| Typical acquisition time (per spectral frame): | 5 to 30 ms |
| Typical acquisition time (per hypercube): | 1 to 5 min |

NIR = Near-infrared; FWHM = Full-width at half maximum; CMOS: Complementary metal-oxide semiconductor; FOV = Field of view; QE = Quantum efficiency.

## 2.2 Samples preparation and data acquisition

HyperProbe1 was used to image a series of fresh surgical biopsies of glioma tissue, to validate its performances in retrieving quantitative information of interest on the composition of samples, as well as to infer pathological characteristics of the tumours akin to what is obtained via traditional histopathological screening. The brain tissue samples involved in the study (an example is shown in Fig. 1f) were obtained from fresh surgical excisions of patients taken during routinely performed neurosurgery for brain tumour resection at the Azienda Ospedaliero-Universitaria Careggi (University Hospital of Florence), in Florence. Authorization for the study (Studio ID: 23672 - 23672_BIO) was granted by the Ethical Committee of the Area Vasta Centro Toscana, under Italian law and regulations. Informed consent was collected from each patient involved in the study.

A total number of 11 samples (n = 11) were imaged and analysed with HyperProbe1, to guarantee a degree of statistical robustness in the reconstructed spectra and to investigate specimen variability. The samples were composed of portions of the same tissue removed during the resection that is normally sent to the laboratory for histopathological screening, which classified the type of the tumour based on WHO gradings[14]. This class of information for all the samples is reported in Table 3, providing a broad characterisation of various typologies of glioma.

**Table 3** Classification of the tissue samples investigated with HyperProbe1.

| Sample identifier | WHO grading | Additional info |
|---|---|---|
| S1 | IV | |
| S2 | IV | |
| S3 | III | Discarded due to fragmented size |
| S4 (FOV1) | IV | Two separate FOVs were acquired on the same sample |
| S4 (FOV2) | IV | Two separate FOVs were acquired on the same sample |
| S5 | IV | Labelled with fluorescein |
| S6 | II | Discarded due to presence of light interference |
| S7 | II | Possibly shifting to higher grade (III) |
| S8 | III | Labelled with fluorescein, presence of coagulated tissue |
| S9 | IV | Labelled with fluorescein, but negative |
| S10 | II | Labelled with fluorescein |
| S11 | IV | Labelled with fluorescein, but negative |

FOV = Field of view.

Two samples were discarded from the analysis: sample S3 was a biopsy composed of very small and dispersed fragments of brain tissue (2-3 mm each at most) and, due to their dimensions, the acquired spectra appeared very flat (we hypothesise a large influence of partial volume effect); conversely, sample S6 presented ambiguous patterns due to potential light interference at some wavelengths. In addition, for sample S4, due to its slightly larger dimensions than the rest of the biopsies (5-6 cm), it was decided to acquire two separate FOVs on the same tissue to further analyse variability within subjects. Finally, samples S8, S9, S10 and S11 were marked with fluorescein during the surgery, albeit S9 and S11 were confirmed to be negative to fluorescent emission.

For the HSI data acquisition, a portion of the glioma samples (average size of 2-3 cm) was preemptively washed in phosphate-buffered saline (PBS) to eliminate blood and other unwanted residuals, then imaged on its surface with HyperProbe1 (79 wavelength bands at 5-nm steps between 510 and 900 nm) within 1 hour after excision. A thin glass coverslip was placed over each sample to flatten its top surface for uniform focusing of the FOV, whilst a dark absorbing material was placed at the bottom to avoid any potential reflection of the light back into the tissue. The acquisition of a single hypercube for every sample was typically less than 5 minutes, a short enough duration to ensure that the tissue had not deteriorated nor oxidised during the imaging.

### *2.2 Data processing and spectral unmixing analysis*

Reflectance hypercubes $R(x, y, \lambda)$ for each sample were reconstructed by normalising the hyperspectral data of the reflected light intensity $I(x, y, \lambda)$ acquired with HyperProbe1 with reference hypercubes $W(x, y, \lambda)$ obtained using a white calibration standard (Labsphere, Spectralon® 5"), after dark counts subtraction $D(x, y, \lambda)$, and by weighting the latter two datasets for the ratios of their corresponding integration times $t$ of the camera used during the acquisition. The formula for the reconstruction of the reflectance hypercubes is

$$R(x,y,\lambda) = \frac{I(x,y,\lambda) - \frac{t_I}{t_D} D(x,y,\lambda)}{\frac{t_I}{t_W} W(x,y,\lambda) - \frac{t_I}{t_D} D(x,y,\lambda)} \quad (1)$$

where $t_I$, $t_D$ and $t_W$ are the camera integration times for each frame of the intensity, dark and white hypercubes, respectively.

A fast, spectral unmixing approach based on modified Beer-Lambert's law (MBLL) was used to infer the differences in the molecular composition of the biopsies, as described in Ezhov *et al*[16]. These differences were quantified with respect to sample S1, using the average spectral reflectance of the central area of the sample as baseline spectrum. Computational time to analyse a full dataset from each biopsy was about 2-3 minutes with two AMD EPYC 7452 32-Core processors.

We then compared the inferred compositions for two different scenarios: (1) by fitting the whole measured wavelength range (from 510 nm to 900 nm), and (2) by fitting only the NIR portion of the available spectrum (in our case, from 740 nm to 900 nm). For the latter, we expected the major absorbing chromophores to be oxygenated ($HbO_2$) and deoxygenated (HHb) haemoglobin, the oxidised (oxCCO) and reduced (redCCO) forms of cytochrome-c-oxidase (CCO), as well as water and lipids[8,17,18]. The spectral signatures of the chromophores targeted by the analysis are depicted in Fig. 2a and Fig. 2b. In the visible range, we also assumed the presence of additional chromophores, specifically the oxidised and reduced forms of cytochrome-b (Cyt-B) and cytochrome-c (Cyt-C), due to their involvement in the metabolic processes[17].

The inferred compositions, in the forms of either concentrations or volumetric contents, are in units [mM/cm] and [$cm^{-1}$], respectively, as we used unitary pathlength (1 cm) in our experiments. We have previously seen that a quasi-constant pathlength only effectively scales the concentrations, at least in the NIR range[16], and is therefore sufficient for the preliminary task of attempting to distinguishing biopsies of different tumour gradings.

*2.3 Monte Carlo simulations of penetration depth in tissue*

A 3D, *in silico*, optical and geometrical model of brain biopsy was designed to assess and quantify the depth of penetration of light in the tissue, at the various spectral bands of HyperProbe1[9]. This was done in MATLAB using a voxel-based Monte Carlo (MC) simulation software. The software chosen was Monte Carlo eXtreme (MCXLAB)[20,21], which simulates photon transport within a 3D, voxel-based model with arbitrary optical properties. The simulations were carried out on a desktop computer with an Intel Xeon W5-3425 and two RTX 4090 GPUs.

The simulated geometry consisted of a homogeneous, semi-infinite slab of grey matter with a thickness of 0.5 cm (taking into account the maximum thickness recorded among the investigated samples). Isotropic voxels were used, with dimensions of 0.05 mm. This was chosen as it offered an acceptable trade-off between accuracy and computational performances. A black absorbing layer of 0.005-cm thickness with a significantly higher absorption coefficient (in the order of $10^6$ $cm^{-1}$) was placed at the bottom of the slab, to represent the absorbing material used below the biopsies. Fresnel reflection was implemented at the top and bottom boundaries of the model[20].

The reduced scattering coefficient $\mu_s'$ of the grey matter of the model was adopted from Jacques *et al*[18], and was varied with wavelength according to the following equation:

$$\mu_s' = a\left(\frac{500\text{ nm}}{\lambda}\right)^{-b}, \tag{2}$$

where $a$ = 40.8 $cm^{-1}$, and $b$ = 3.089. From $\mu_s'$, the scattering coefficient $\mu_s$ was calculated as an input for MCXLAB with the equation:

$$\mu_s = \frac{\mu_s'}{1-g}, \tag{3}$$

where $g$ is the anisotropy factor of grey matter standing at 0.85 and chosen to be constant, as its variation with wavelength has been demonstrated to be minimal for cerebral tissue[22]. Furthermore, the refractive index of the biopsy model was set to 1.36[18].

Finally, the 3D model of brain biopsy was assumed to be composed of the most absorbing and scattering tissue chromophores, i.e., water, lipids, $HbO_2$, HHb, oxCCO and redCCO[9]. Thus, the total absorption coefficient $\mu_a$ of the model was determined using the equation[9,18]:

$$\mu_a = W \cdot \mu_{a,H_2O} + F \cdot \mu_{a,fat} + ln10 \cdot C_{HHb} \cdot \varepsilon_{HHb} + ln10 \cdot C_{HbO_2} \cdot \varepsilon_{HbO_2} + \\ + ln10 \cdot C_{oxCCO} \cdot \varepsilon_{oxCCO} + ln10 \cdot C_{redCCO} \cdot \varepsilon_{redCCO}, \quad (4)$$

where $W$ and $F$ are the water and lipids volumetric contents, respectively; $\mu_{a,H_2O}$ and $\mu_{a,fat}$ are the absorption coefficients of water and lipid, respectively; $C_{HHb}$, $C_{HbO_2}$, $C_{oxCCO}$, and $C_{redCCO}$ are the molar concentrations of HHb, $HbO_2$, oxCCO and redCCO, respectively; and $\varepsilon_{HHb}$, $\varepsilon_{HbO_2}$, $\varepsilon_{oxCCO}$ and $\varepsilon_{redCCO}$ are the molar extinction coefficients of HHb, $HbO_2$, oxCCO and redCCO, respectively. $W$, $F$, $C_{HHb}$, $C_{HbO_2}$, $C_{oxCCO}$, and $C_{redCCO}$ are derived from Giannoni *et al*[9], whereas the absorption coefficients $\mu_{a,H_2O}$ and $\mu_{a,fat}$ (graphed in Fig. 2a), as well as the molar extinction coefficients $\varepsilon_{HHb}$, $\varepsilon_{HbO_2}$, $\varepsilon_{oxCCO}$ and $\varepsilon_{redCCO}$ (graphed in Fig. 2b) were derived from Giannoni *et al*[9] and Prahl *et al*[19]. Table 4 summarises the main composition of the 3D *in silico* model of brain biopsy used for the MC simulations, based on human grey matter.

**Table 4** Composition of the 3D *in silico* model of brain biopsy used for the MC simulations.

| Model composition | Values |
|---|---|
| Water content, $W$ | 70% |
| Lipid content, $F$ | 10% |
| Molar concentration of HHb, $C_{HHb}$ | 56.7 μM |
| Molar concentration of $HbO_2$, $C_{HbO_2}$ | 56.7 μM |
| Molar concentration of oxCCO, $C_{oxCCO}$ | 1 μM |
| Molar concentration of redCCO, $C_{redCCO}$ | 4 μM |

HHb = Deoxygenated haemoglobin; $HbO_2$ = Oxygenated haemoglobin; oxCCO = oxidised cytochrome-c-oxidase; redCCO = reduced cytochrome-c-oxidase.

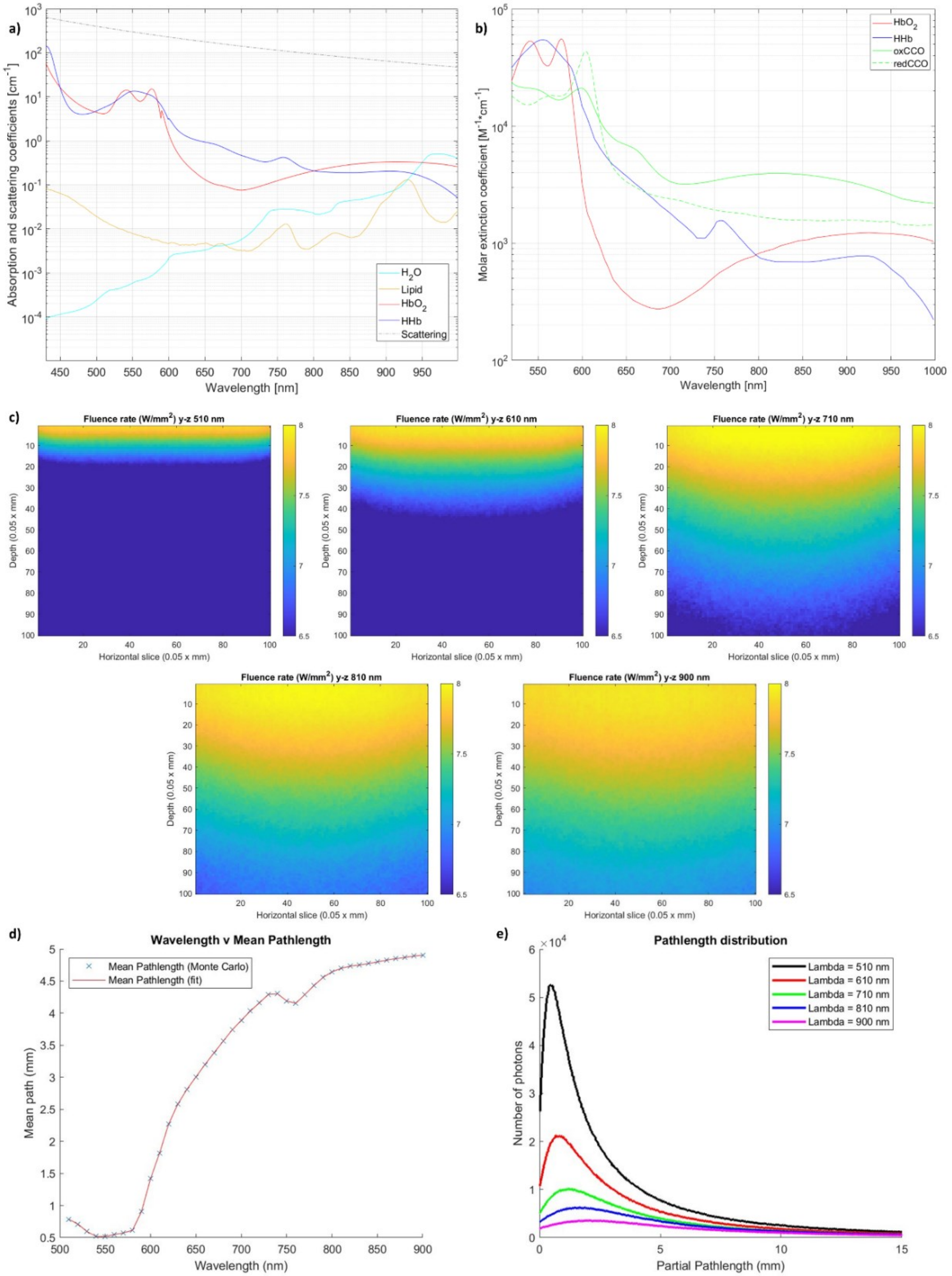

**Fig. 2** (a) Absorption coefficients for $HbO_2$, HHb, lipids and water, and scattering coefficient of generic brain tissue in the visible and NIR range (150 g/L concentration of Hb in blood and blood volume content in generic brain tissue equal to 5% are assumed)[8,19]; b) Molar extinction coefficients of $HbO_2$, HHb, oxCCO and redCCO in the visible and NIR range[8,17]; c) Simulated fluence rates within the 3D cerebral biopsy model at different wavelengths (the plots are in a logarithmic scale and the absorbing layer was excluded from the plots to maintain a visible contrast). The distribution of the fluence showed a noticeably higher penetration of the light at the NIR wavelengths; d) Simulated mean photon pathlength within the 3D biopsy model, as a function of wavelength; e) Partial pathlength distributions of the photons simulated within the 3D biopsy model at various wavelengths (as previously, the absorbing layer was not included in this figure due to a negligible number of photons passing through it).

A planar, divergent light source was positioned above the 3D biopsy model to simulate the illumination of the grey matter slab at the same wavelengths of operation of the HyperProbe1 (500-900 nm, at 5-nm sampling). The direction of the illumination was normal the z axis. A total of $10^7$ photons were simulated for each wavelength band, with emission power equal to 1 mW.

Simulated fluence rates for each wavelength were obtained from the MC simulations, allowing to visualise and assess the spatial distribution of the photons at the different spectral bands within the 3D model. Examples of these distributions are depicted in Fig. 2c, demonstrating increasing depth of penetration of the light within the tissue for longer and longer wavelengths. The depth of penetration of the photons in the model of biopsy ranged from 10-15% of its thickness (about 0.5-0.75 mm) for the visible light, up to almost the whole size of the sample (5 mm), for the NIR light.

To further investigate the relationship between wavelength and penetration depth of the light, the mean pathlength across the tissue of the simulated photons was estimated from the MC simulations for each spectral band, as reported in Fig. 2d. The results further demonstrate that the mean pathlength travelled by the photon within the tissue increases as the wavelength of the light gets longer, up to about 5 mm (at 900 nm). Nonetheless, this gradual increase is not homogeneously continuous, as local extrema are identifiable in Fig 2c, corresponding to the peaks of absorption of haemoglobin at about 530-570 nm and 770 nm (as visible in Fig. 2a).

Finally, distributions of the partial pathlengths of the simulated photons were estimated for each wavelength of HyperProbe1: Fig. 2e depicts examples of these distributions across the full range of the system (the absorbing layer at the bottom is excluded). The results, together with the outcomes previously illustrated, demonstrate empirically that HyperProbe1 does not simply map the optical properties of the investigated tissue on its visible surface, but actually reconstruct an integrated distribution on 2D of the spectral features of the biopsy samples across their entire 3D volume, thanks to the use of both visible and NIR light with different penetration capabilities.

# 3 Results

## *3.1 Qualitative and comparative evaluation of the HyperProbe1 data*

Preliminary qualitative evaluation of the data collected with HyperProbe1 on the cerebral *ex vivo* tissue was conducted, to assess both heterogeneity in the spectra within the same sample, as well as spectral variability between all the biopsies. For intercomparison across the various samples described in Table 3, each processed reflectance hypercube $R(x, y, \lambda)_n$, for $n$ = S1…S11, was averaged spatially over its entire FOV, in order to obtain a single averaged reflectance spectrum for each biopsies. All these reflectance spectra are shown altogether in Fig. 3a. Overall, the average reflectance spectra between samples share similar trends, reporting low reflectance in the visible range (where absorption from haemoglobin is at its highest, as per Fig. 2a), which then tends to increase gradually towards the NIR range beyond 600 nm, where scattering becomes predominant. However, significant variability in both magnitudes of the spectra in the same regions and in their local features are also present, differentiating the signatures of the various samples. Such aspect is further highlighted by averaging the abovementioned spectra according to their WHO grading: lower grade glioma (LGG) samples (WHO grade II and III) were grouped together and the mean

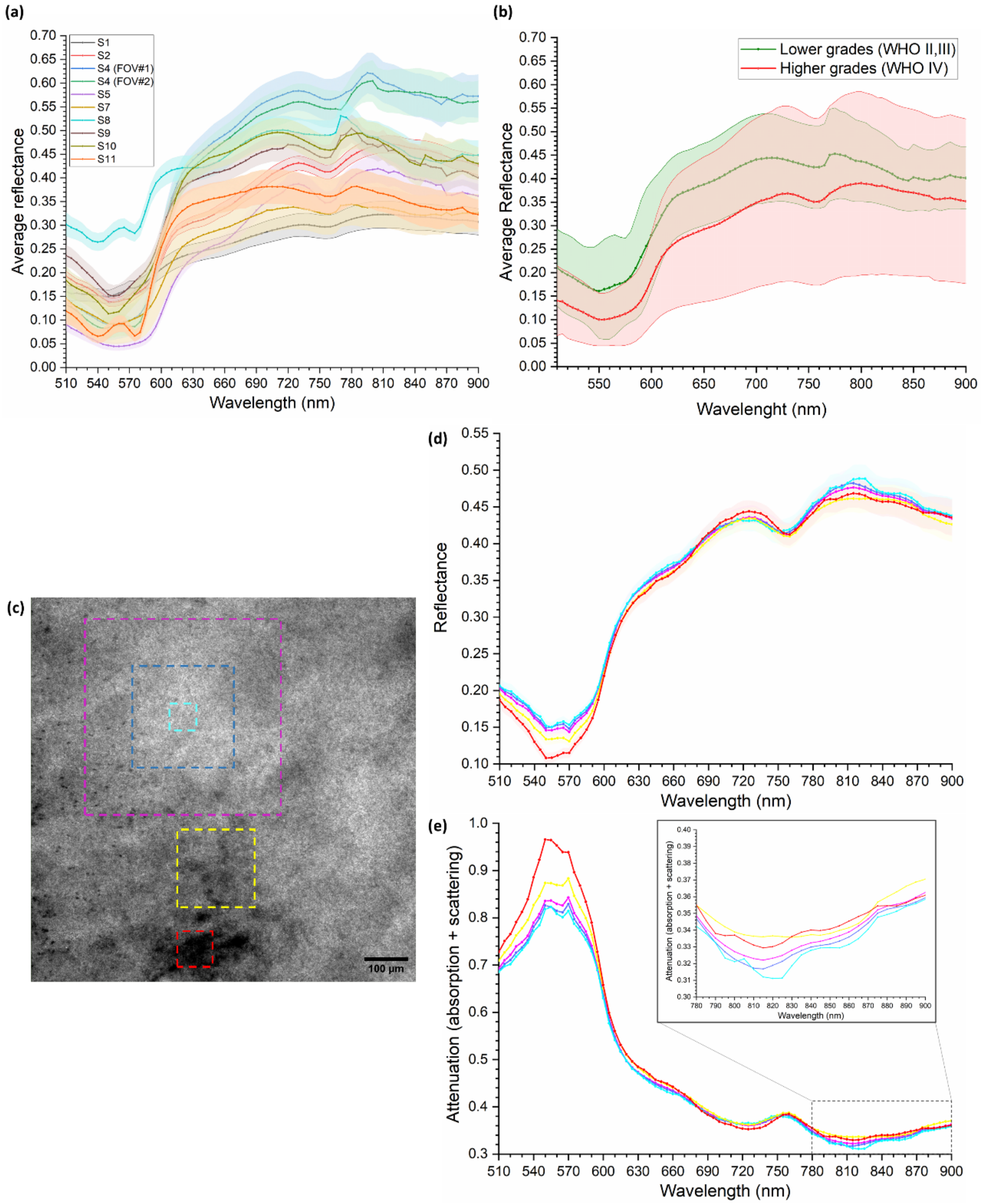

**Fig. 3** (a) Intercomparison of averaged reflectance spectra over the entire imaged FOVs of each biopsy sample; (b) Comparison between average reflectance spectra grouping LGG (WHO II, III) against HGG samples (WHO IV); (c) Processed spectral image from HyperProbe1, at 560 nm, of HGG (WHO grade IV) biopsy sample S2, with highlighted, selected ROIs in the FOV in which average reflectance spectra were calculated; (d) Example of average reflectance spectra in the corresponding ROIs of the biopsy sample S2; (e) Example of average attenuation spectra in different ROIs of biopsy sample S2, with the portion within the NIR range 780-900 nm enlarged.

of their average reflectance spectra across the whole FOVs was compared against the same mean for all the higher grade glioma (HGG) samples (WHO grade IV), as reported in Fig. 3b. It can be seen that there are indeed significant differences in the means between LGG and HGG samples, regarding both local spectral trends as well as magnitude, with the LGG mean presenting higher reflectance on average along the entire spectral range. Conversely, HGG shows larger variance among biopsies of their corresponding group, for the same grade.

A number of different regions of interest (ROI) were selected on the reflectance hypercubes $R(x, y, \lambda)$ to calculate average reflectance spectra, including different sizes and portions of the FOV reporting identifiable visual features, such as blood clusters. This was done to investigate potential spatial changes in the reflectance spectra across the FOV of the samples. Fig. 3c shows an example of a single, processed reflectance spectral image (as described in Sec. 2.2) collected with HyperProbe1 for arbitrary HGG (WHO grade IV) sample S2, at the bandwidth centred at 560 nm, whereas Fig. 3d depicts the corresponding average reflectance spectra. The comparison of the reflectance spectra for each sample on different ROIs highlights significant (from 0.0017 to 0.0378 average root mean square deviation (RSMD) for the reflectance curves across all samples) differences in their shapes and trends, as visible for the model case of S2 in Fig. 3d, for two specific spectral ranges: (1) between 510 and 660 nm, and (2) between 780 and 880 nm. In contrast, outside of these ranges, the remainder of the spectral signatures displays homogeneous distribution of the optical properties of the *ex vivo* samples across the entire FOVs. The largest differences (0.0378 of average RSMD across samples) are reported for the ROIs where accumulation of blood are clearly visible. For the first mentioned range (510-600 nm), such results could be correlated to the presence and strong influence of the visible peaks of haemoglobin absorption (Fig. 2a). The influence of the absorption of haemoglobin could also be connected to the reported differences in the spectra for the second mentioned range (780-880 nm), which is characterised by a broad peak of absorption from $HbO_2$. However, differences are identified in the biopsy samples also for ROIs not including visible blood clusters: this could then be connected to local differences in the concentrations of CCO, as the identified range overlaps with the NIR absorption peak of the latter (Fig. 2b), as well as for the increasing weight of the absorption of water and lipids towards the end of the NIR range (Fig. 2a).

For a more direct comparison between the spectral signatures of the biopsies reconstructed with HyperProbe1 and the pure optical signatures of the chromophores of interest reported in Fig. 2a and Fig. 2b, we calculated the attenuation spectra $A(x, y, \lambda)$ associated with the contribution of optical absorption and scattering in the tissues, from the reflectance spectra $R(x, y, \lambda)$ in the same ROIs of the samples, using the formula: $A(x, y, \lambda) = -log_{10}\big(R(x, y, \lambda)\big)$. Fig. 3e reports, for instance, the average attenuation spectra of the same HGG sample S2 and the previously selected ROIs. In the range 510-600 nm, the attenuation spectra of the brain tissue samples correlate with the combined profiles of the absorption spectra of $HbO_2$ and HHb, with noticeable peaks at around 545-555 nm and at 575 nm. Another peak is also identified in all samples at around 755-760 nm, overlapping the equivalent one from the absorption spectra of HHb. In all samples, attenuation in the range 510-600 nm is reported to increase gradually when shifting from ROIs with no visible accumulations of blood towards ROIs that include the latter at different degrees of covering (as visible for the case of S2 in Fig. 3d). Figure 3e highlights and enhances the visualisation of the attenuation spectra from S2 in the range between 780 and 900 nm, where differences are reported for all the ROIs regardless of the presence of any discernible spatial feature or difference in contrast. This is occurring largely across all the analysed samples of surgical biopsies. In particular, localised peaks of attenuation corresponding to about 840 nm are identified in a number of ROIs in the samples, where the optical absorption of oxCCO is also at its highest. The reported discrepancies among concentric ROIs of different sizes in relatively homogenous areas of the *ex vivo* tissue could be linked to either local variation in the abovementioned chromophore, as suggested by the overlapping of the peaks, or to partial volume effects connected to changes in the optical pathlength travelled by the lights within the sampled regions[9].

### *3.2 Quantitative biochemical analysis of the HyperProbe1 data with spectral unmixing*

As mentioned in Sec. 2.2, we compared the inferred compositions of the expected chromophores from the spectral unmixing algorithm in all the biopsy sample, for two different fitting scenarios: in the whole range from 510 nm to 900 nm, and in the NIR portion of the spectrum from 740 nm to 900 nm. For the first scenario, we obtained satisfactory spectral fits matching the measured attenuation: an example is depicted in Fig. 4a and Fig. 4b, for HGG sample S4 in FOV#1 (WHO grade IV) and LGG sample S10 (WHO grade II), showing also the reconstructed quantitative maps for the total concentration of haemoglobin (HbT), given as the sum of $HbO_2$ and HHb, as well as for the concentration of differential CCO (diffCCO), given as the difference between the concentrations of oxCCO and redCCO[8,17]. Blood clusters are resolved with high resolution, due to the haemoglobin peaks in the 500-600 nm range and the expected high concentration and absorbance of haemoglobin (compared to the other known chromophores). Similar well-matching spectral fits via the MBLL are found across all patients, as we report the root mean square error (RMSE) means across all pixels and across all patients to be in the range 0.017 to 0.039, which is of similar magnitude to the exemplary RMSEs reported in Fig. 4.

A preliminary attempt to classification of the biopsy samples from the inferred hyperspectral results was also performed. As shown in Fig. 4c and Fig. 4d, we found that predicted lipids content allows to separate LGG (grade II and III) biopsies from HGG biopsies (grade IV). Even though we observe significant overlap between the different distributions (with an overlap coefficient of 49.96% assuming normal distributions), we generally noticed a trend of higher differential lipids content in HGG biopsies, with a mean of -0.466 $cm^{-1}$. Conversely, the LGG samples were found to have a mean lipid content of -2.53 $cm^{-1}$. The distinction between the two gradings for all samples was possible by computing the means across all the biopsies and using the -2 $cm^{-1}$ lipid content difference threshold, as shown in Fig. 4d. This difference in means of the tumour grades was also found to be statistically significant via the Mann-Whitney U test ($p=0.017$), testing for equal means. Conversely, we did not find evidences of CCO or haemoglobin (established biomarkers for cellular metabolism and haemodynamics, respectively[8]) to be able to separate gradings of glioma samples in the whole range 510-900 nm (as it can be seen in Fig. 4d, for CCO). The overlap coefficient between LGG and HGG samples was found to be considerably larger, with 81.2% and 84.6% for inferred diffCCO and $HbO_2$ concentrations, and statistical differences in grading were not observed ($p=0.84$ and $p=0.99$, respectively). On a final note, as seen on Fig. 4d, by considering both variables (lipid content and diffCCO concentration) on the 2D distribution plot, an oblique line could arguably even better separate the two glioma grades. A larger sample size will be required to test such more complex hypotheses further.

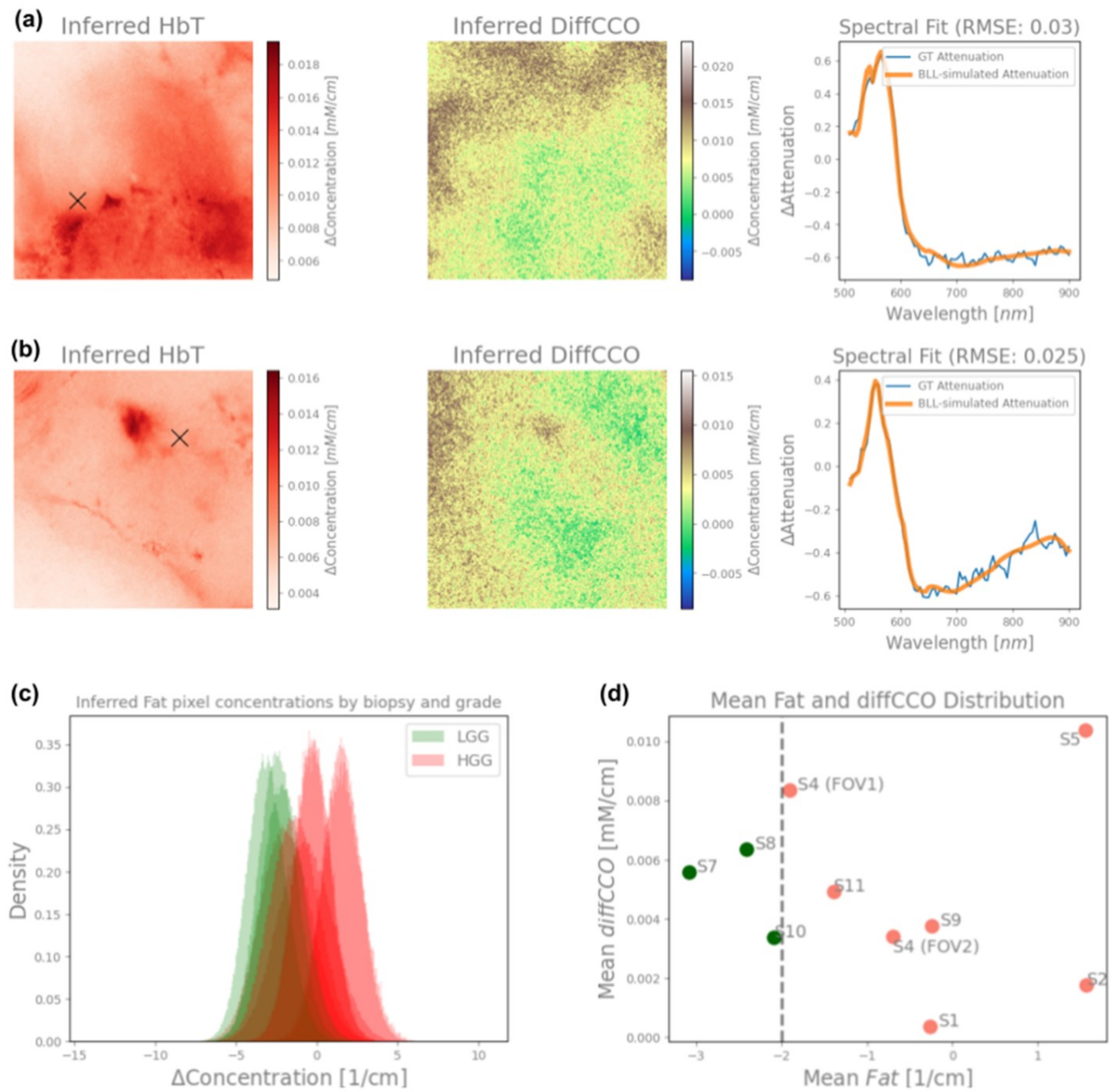

**Fig. 4** Inferred HbT and diffCCO concentration maps of HGG S4 FOV#1 (a) and LGG grade S10 (b) samples fitting the whole measured wavelength spectrum (510-900 nm), whilst a model fitting of the observed attenuation for the marked pixel in the HbT image is shown in the third column for both biopsies; (c) Histogram showing probability density distribution of inferred lipid volumetric content of each pixel across different LGG (displayed in green) and HGG (displayed in red) grade samples; (d) Distribution means of the lipid content (reported on x-axis) and diffCCO concentrations (reported on y-axis) suggest that lipid mean content could be able to distinguish grading of all samples, whereas no apparent separation is visible for the inferred mean diffCCO concentrations.

In the second scenario, we estimated inferred compositions within the biopsy samples using exclusively the NIR range between 740 nm and 900 nm, which was chosen to target chromophores that are known for their characteristic absorption profiles in such range, particularly oxCCO and redCCO (as seen in Fig. 2b). Indeed, MBLL has been commonly employed in this specific NIR range to infer differences in metabolic activity using CCO as a biomarker[8–10,17]. As shown in Fig.5a and Fig. 5b, we again fit the observed signal qualitatively well: the inter-biopsy mean RMSE errors are found to be in the range 0.0151 to 0.296, i.e., the spectral fits of all biopsies can be expected to be similar as observed in Fig.5a and Fig. 5b, as for the previous scenario, albeit being slightly worse due to the expected reduction in SNR at the latter end of the measured spectrum, where scattering of light become predominant. Notable loss in the resolution of various blood clusters can be observed, which was also expected due to the exclusion of the 510-600 nm range with larger haemoglobin absorption peaks. Interestingly, we observed significant differences to the inferred molecular concentrations across biopsies of LGG and HHG that do not match the ones highlighted by the spectral analysis on the whole wavelength range: e.g., in this scenario, the inferred lipid contents were not able to distinguish between the different grades of the samples. We reported an overlap coefficient of 77.1% and no statistical differences in the concentration means for lipids in this scenario ($p=0.84$). However, we observed average concentrations differences of metabolic diffCCO between LGG and HGG samples, with HGG samples showing lower diffCCO mean concentrations, as seen in Fig. 5c and Fig. 5d. The diffCCO concentration threshold at 0.35 mM/cm was able to distinguish all LGG and HGG samples by computing the intra-biopsy diffCCO concentration mean ($p=0.017$), despite visible overlap (with overlap coefficient of 61%) between the distributions as seen in Fig. 5c. Singularly, we also reported that both oxCCO and redCCO individually are seemingly not able to distinguish biopsy grading, as we find overlap coefficients of 84.6% and 73.3%, and no statistically significant differences ($p=0.99$ and $p=0.12$, respectively). This is shown in Fig. 6 plotting probability density distributions of each pixel across all biopsies, divided between LGG and HGG. Only the difference diffCCO between the two inferred concentrations resulted in the suggested distinction of the two classes of tumoral grades.

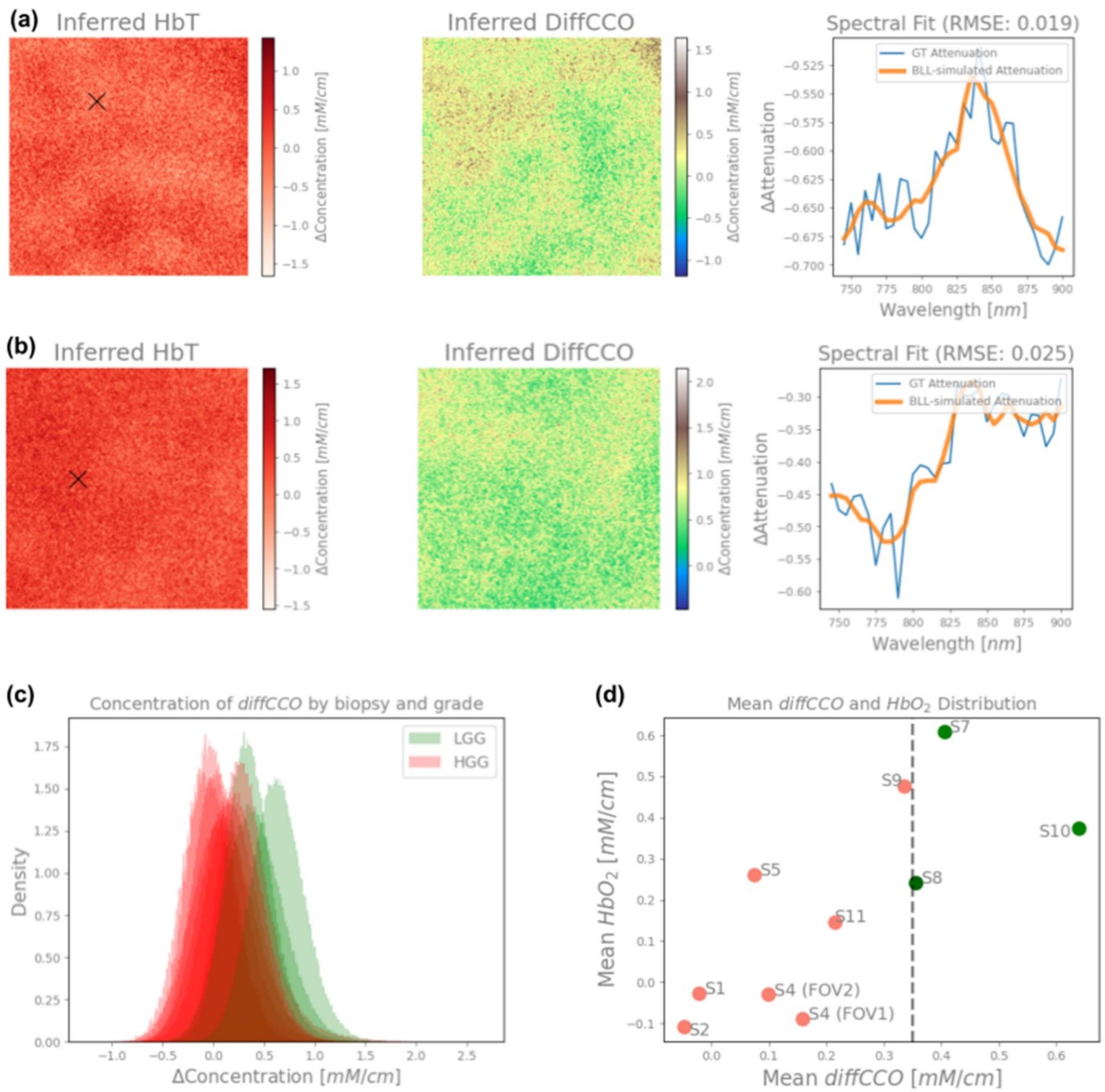

**Fig. 5** Inferred HbT and diffCCO concentration maps of HGG S4 FOV#1 (a) and LGG S10 (b) samples fitting exclusively the NIR range (740-900 nm), whilst a model fitting of the observed attenuation for the marked pixel in the HbT image is shown in the third column for both biopsies; (c) Histogram showing probability density distribution of inferred diffCCO concentrations of each pixel across different LGG (displayed in green) and HGG (displayed in red) grade samples; (d) Distribution means of diffCCO and $HbO_2$ concentrations suggest that these parameters could be able together to distinguish lower and higher grade samples of glioma tissue, with diffCCO being the most accurate, across all the investigated samples.

Furthermore, as depicted in Fig. 5d, we also found an approximately linear correlation in the 2D domain between the mean differences in concentration of diffCCO and those of $HbO_2$: together with a reduced DiffCCO mean concentration, HGG samples also present a correlated reduction in the mean concentration of $HbO_2$, with the combination of the two biomarkers enhancing the possibility of separating samples grading more effectively.

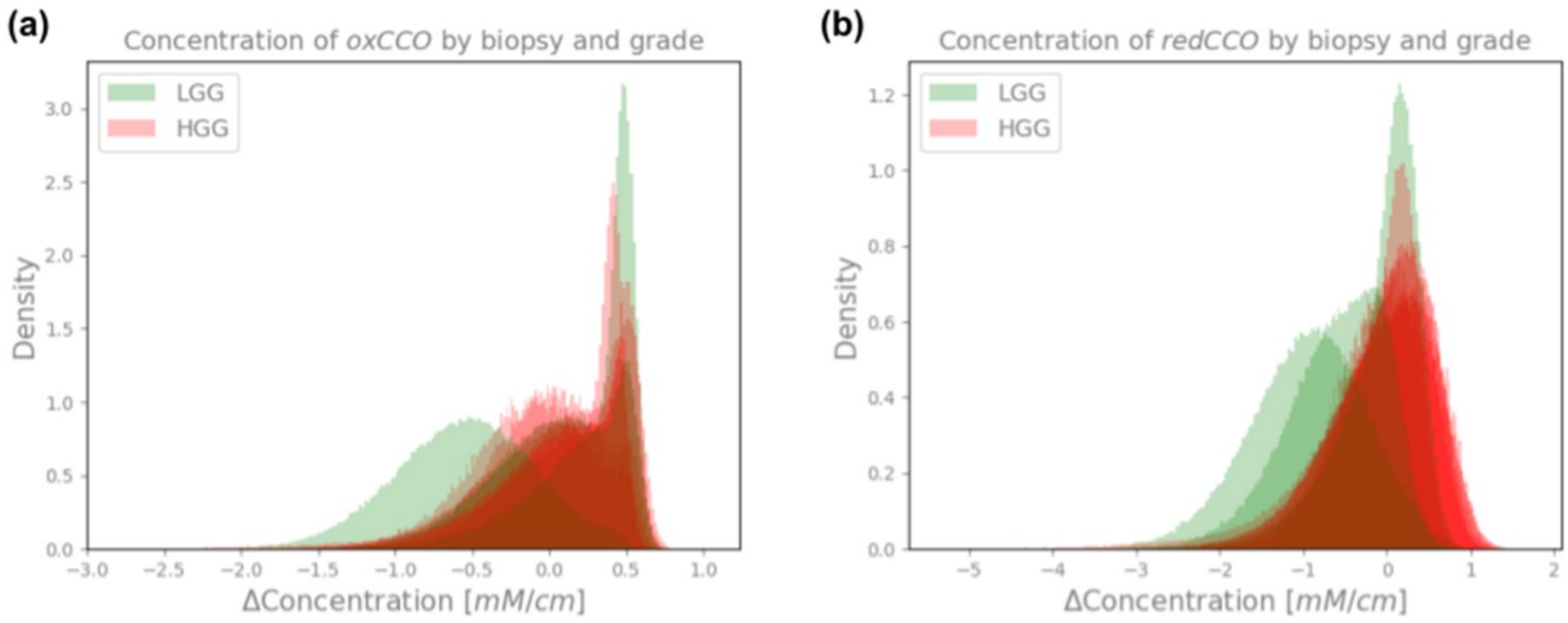

**Fig. 6** Histogram showing probability density distribution of inferred oxCCO (a) and redCCO (b) concentrations of each pixel across different LGG (displayed in green) and HGG (displayed in red) grade samples**.** Neither the inferred oxCCO and redCCO density distributions using the NIR range (740-900 nm) suggest to be able to differentiate tumour grading, contrary to the diffCCO mean concentrations.

On a final note, we observed that the HGG sample S9 additionally displays diffCCO characteristics close to LGG samples (Fig. 5d): we hypothesise that there could be different (possibly concurrent) reasons for this low distinguishability: even though it has not been shown that HGG samples can show characteristics of LGG samples, the opposite (i.e., high grade-typical histological characteristics in low grade lesions) has recently been reported by Motomura *et al*[23]. Thus, in general, we suggest that it might be possible that lower grade lesions could display characteristics of higher grade lesions, although further investigation on this would be needed.

## 4 Conclusions

We presented a novel, transportable HSI device, name HyperProbe1, based on fast spectral scanning via a combination of SCL illumination and AOTF filtering. At full operational performances,

HyperProbe1 can scan sequentially up to 79 wavelengths between 510 and 900 nm, with 5-nm sampling and high spectral resolution (3.5-7 nm of bandwidth), as well as acquire corresponding spectral images on a 0.9 x 0.9-$cm^2$ FOV at high spatial resolution (4.38 μm), in less than 5 minutes.

Furthermore, we also provided a preliminary assessment of the data acquired and analysed with HyperProbe1 on a number (n = 11) of fresh surgical samples of glioma from patients at different WHO gradings, including both lower grade (WHO grades II and III) and higher grades (WHO grade IV) tumours. The initial findings demonstrated the capability of the device to reconstruct quantitative maps of the distribution of the concentrations of various chromophores of interest, in particular haemoglobin and CCO (both established biomarkers for tissue haemodynamics and metabolism, respectively), as well as lipids volumetric content. We also looked at differences in the inferred contributions of these biomolecules between LGG and HGG biopsies, that may suggest a way to distinguish between the two, and thus potentially provide the basis for a HSI-based methodology for tumour classification. We found indeed significant differences between mean lipid contents across samples that could potentially be used to distinguish tumour grading, when fitting over the entire work range (510-900 nm). In particular, HGGs presented higher mean lipid content than LGGs: a plausible biological interpretation of this could be connected to a substantial lipid storage for lipid metabolism in higher grade gliomas, a known characteristic feature of GBM[24,25].

Similarly, we also found that the analysed HGGs presented inferior concentrations of diffCCO compared to LGGs, when fitting the data exclusively in the NIR range (740-900 nm): this may suggest a potential metabolic path with NIR light to distinguish lower and higher grade samples. This finding may also emphasise the potential role of diffCCO in providing metabolic differences that lies beyond its critical use in NIR spectroscopy (NIRS) applications[17]. The exact mechanisms by which these concentration differences in diffCCO arise still remain uncertain: as changes in the concentration of diffCCO are directly proportional to variations in the metabolic activity of the tissue, one would expect an influence of enhanced hypermetabolism in tumours at higher grades. However, HGGs (such as GBMs) are characterised by the unique presence of necrotic tissue as the core part of the lesion, with the hypermetabolic region occurring only at the borders[26,27]. Such central necrotic areas, composed almost entirely of dead -i.e., ametabolic- cells could explain the overall reduced mean concentration of diffCCO in HGG compared to LGG, which instead do not present necrosis. Furthermore, the results showed a direct correlation between the reduction of diffCCO in HGG with a corresponding decrease in $HbO_2$, which could also be connected to the hypoxic microenvironment that is specific of this type of brain tumours[28,29]. From a biological perspective, HGGs, such as GBMs, are known to be triggered and driven in their proliferation by hypoxic microenvironments within the cerebral tissue[28,29]. Such phenomenon could explain the reduced mean concentration of $HbO_2$, which is a biomarker for oxygenation of the brain, and its correlation with the diffCCO, since a direct association between oxygen delivery and its metabolic consumption has been strongly established[30].

Future investigations on a larger and enriched cohort of biopsy samples of various types will be needed to further confirm all these results on a more robust statistical basis. In particular, comparisons with control samples composed of healthy cerebral tissue will also be strongly required for increasing the accuracy and reliability of any prediction.

The goal of HyperProbe1 is to provide a first proof-of-concept application to rapid and quantitative digital histology of *ex vivo* tissue from excised surgical biopsies, in particular of cerebral glioma, by reconstructing quantitative maps of the distributions of chromophores of interest in the tissue via fast spectral unmixing algorithms. In this perspective, we demonstrated that HyperProbe1 can collect and analyse full-range HSI data of light reflectance from surgical biopsies in less than 1 hour after their excision (including preparation of the sample and setting up of the measurements), a significantly much faster time than traditional H&E histopathological screening, which normally takes several days up to few weeks. Such result qualifies HyperProbe1 for an envisioned, future utilisation providing *in situ*, streamlined, and non-destructive screening of fresh tissue samples, ideally within or just next to the operating theatre. This application could considerably benefit the outcome of the surgical treatment and provide an all-optical advancement to current post-surgical histopathological practice.

From a technological perspective, we also aim and improving the current setup of HyperProbe1 to further improve its performances: we are currently working on extending the operational spectral range of the device at both ends, by (1) covering the rest of the available visible range below 500 nm and including also part of the near ultraviolet (UV), as well as by (2) expanding the NIR coverage beyond 900 nm. This approach could further enrich the collected spectral data and allow us to potentially target even additional chromophores of interest[8].

The versatility of HyperProbe1 could also pave the way for potential applications of the instrumentation to virtually any types of surgical and non-surgical biopsy (or other *ex vivo* tissues), as well as even move beyond digital histopathology. In this perspective, HyperProbe1 can also be envisioned as an investigative, high-performances HSI device for preclinical *in vivo* applications: it could be used to explore and tailor features of HSI (such as type and number of wavelengths) that could be translated into clinical settings by specifically engineering more compact, cost-effective and user-friendly medical devices. Within the framework of the HyperProbe project and consortium[31,32], we indeed aim at translating this technology for its use as a new neuronavigation tool during brain surgery, such as in glioma resection. Such a device would aim at providing an innovative approach to guided-neurosurgery, by transforming current practice towards an all-optical, real-time, quantitative and accurate imaging approach, that could significantly help neurosurgeons, enhance the efficacy of the treatment, and ultimately improve life expectancy of the patients.

*Disclosures*

The authors declare no financial conflict of interest.

*Code, Data, and Materials*

In support of open science, the data presented in this article are publicly available on Zenodo at https://zenodo.org/records/10908359. Similarly, all our spectral unmixing code is available on GitHub at https://github.com/HyperProbe/SpectraFit.

*Acknowledgments*

The HyperProbe consortium and project has received funding from the European Union's Horizon Europe research and innovation program under grant agreement No 101071040 – Project HyperProbe. Views and opinions expressed are however those of the author(s) only and do not necessarily reflect those of the European Union. Neither the European Union nor the granting authority can be held responsible for them. AA, FL and IL from UCL are supported by UK Research and Innovation (UKRI) grant No. 10048387

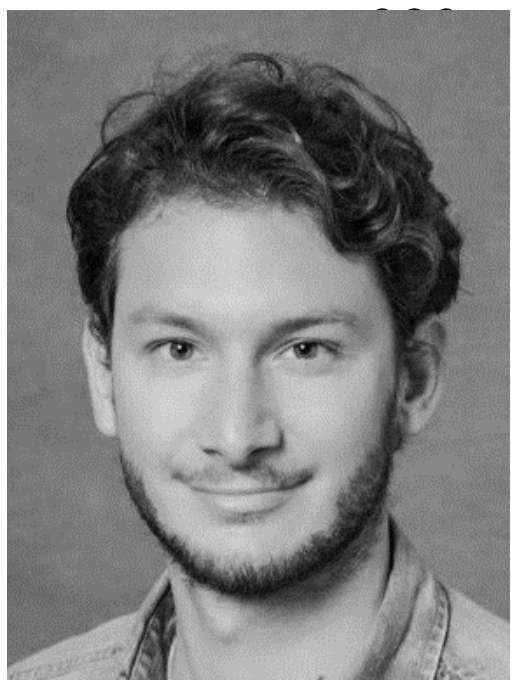

**Luca Giannoni** is a postdoctoral researcher at the European Laboratory for Non-Linear Spectroscopy (LENS) of the University of Florence. He received his PhD in Medical Imaging from University College London (UCL) in 2020. His current research interests focus on developing cutting-edge optical microscopy systems for neuroimaging, as well as on designing and validating cost-effective, compact HSI devices for clinical translation. He is a member of SPIE.

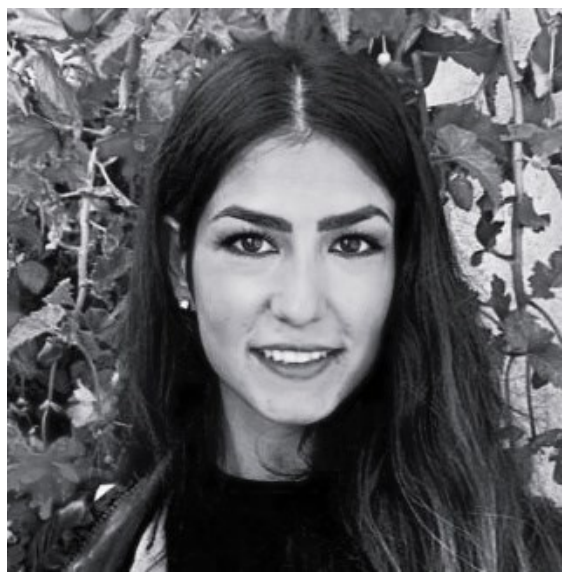

**Marta Marradi** is a PhD student at the European Laboratory for Non-Linear Spectroscopy (LENS) of the University of Florence. Her work specifically focuses on the development of novel imaging systems based on both multispectral and hyperspectral techniques for biomedical applications, concentrating in the dermatologic and neurological fields.

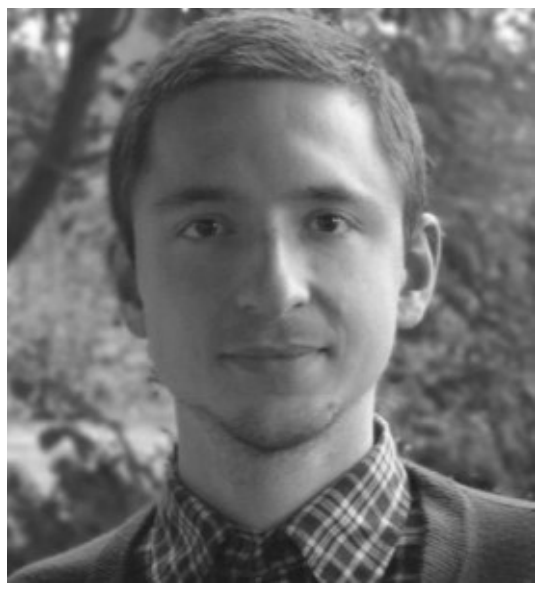

**Ivan Ezhov** is a research scientist at the Institute of Artificial Intelligence in Medicine at the Technical University of Munich. His research focus is at the intersection of biophysical modelling, inverse problems, and deep learning. His interests include computational oncology, physics-based machine learning and spectral unmixing of large scale datasets, such as medical hyperspectral images.

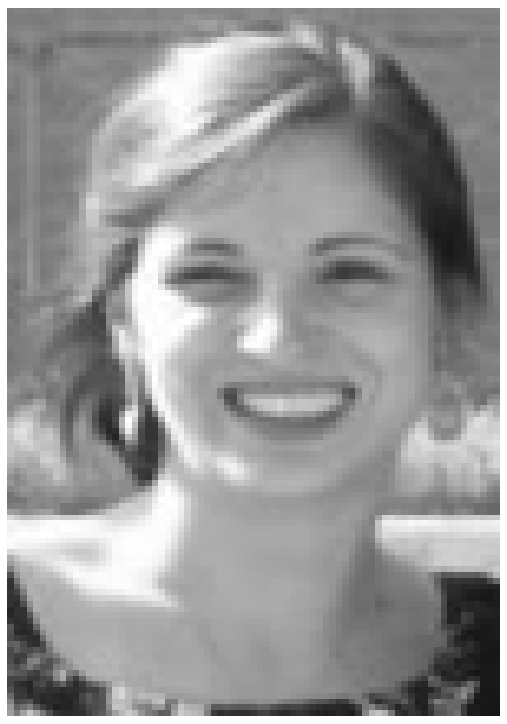

**Camilla Bonaudo** is a neurosurgeon at the Azienda Ospedaliera of Careggi, and a PhD student in Neuroscience at the University of Florence. Her work focuses on the development of new protocols for the application of technologies, such as Navigated Transcranial Magnetic Stimulation (nTMS), to study cognitive functions, acquiring data about brain plasticity and functional reshaping. Her aim is to study brain connectome and comparing functional mapping with the gold standard of the Direct Cortical Simulation.

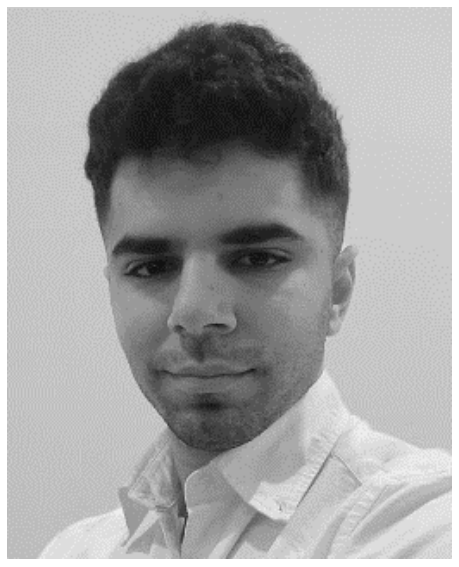

**Angelos Artemiou** is an MRes student at University College London (UCL), specializing in the field of Medical Physics and Biomedical Engineering. His research is primarily focused on developing novel phantoms for the metrological characterisation of NIR and hyperspectral instrumentation.

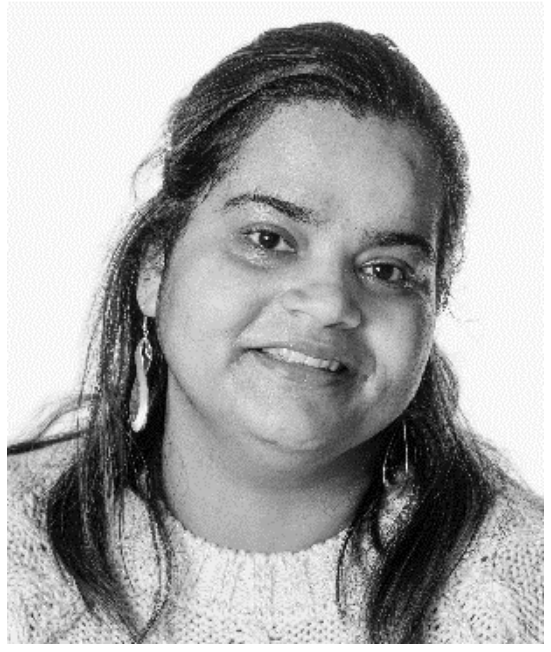

**Anam Toaha** is a PhD student at the European Laboratory for Non-Linear Spectroscopy (LENS) of the University of Florence. She conducts research in the field of biomedical imaging and spectroscopy. She is also part of the HyperProbe project, working on the development of a novel optical imaging device based on hyperspectral imaging, to advance brain surgery by presenting neurosurgeons with enhanced information intraoperatively.

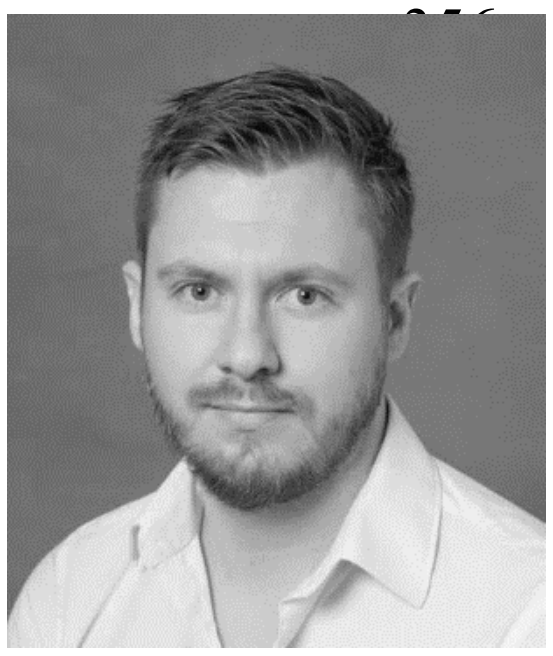

**Frédéric Lange** received his PhD from the University of Lyon and INSA de LYON in 2016. He is now a senior research associate with the Biomedical Optics Research Laboratory, Department of Medical Physics and Biomedical Engineering at University College London. His current main research interests are in the development of novel optical technologies to monitor tissue's oxygenation and metabolism, with a specific interest for non-invasive brain monitoring in healthy (i.e., brain development/neuroscience) and pathological conditions.

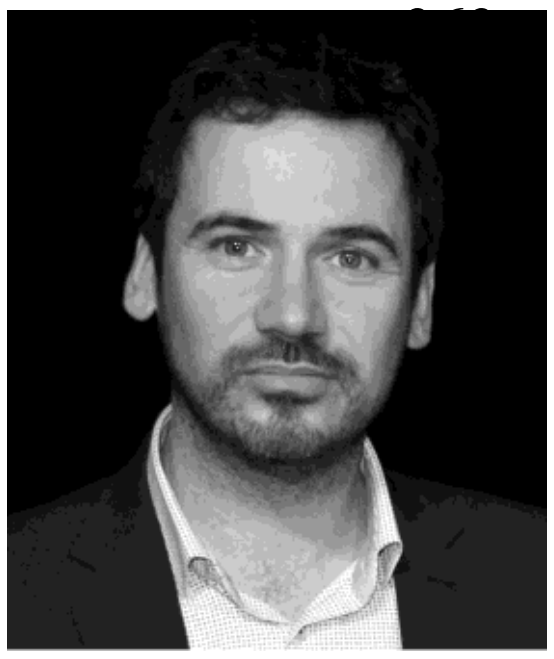

**Bruno Montcel** is a professor at University Claude Bernard Lyon 1. He leads research focused on optical medical devices at CREATIS laboratory and is the chairman of the Biomedical Engineering Department of Polytech Lyon. Its research explores optical imaging methods and experimental set up for the exploration of tissue physiology and pathologies. It mainly fo-cuses on intraoperative and point of care hyperspectral optical imaging methods for medical diagnosis and gesture assistance.

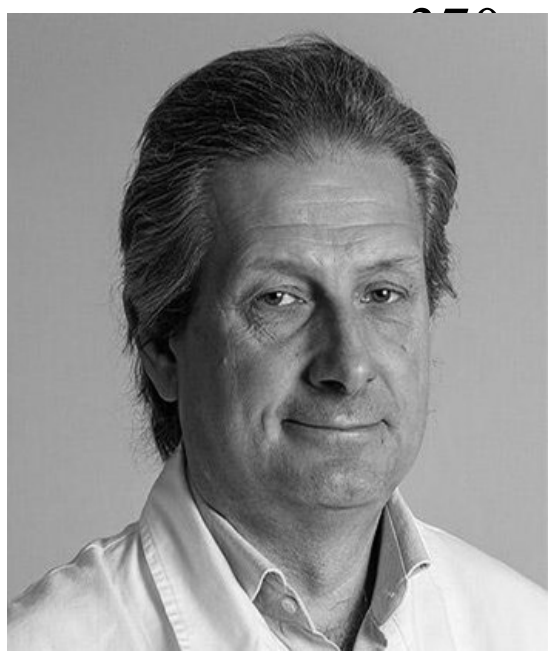

**Alessandro Della Puppa** is a professor of neurosurgery, a neurosurgeon, and the head of the Neurosurgical Department at the Azienda Ospedaliera of Careggi, in Florence. His major interests concern the surgical improvement of neuro-oncology, in particular regarding the development of new intraoperative strategies in awake surgery, in immuno-neuro-oncology, and in navigational TMS in the preoperative and intraoperative mapping of language and cognitive functions, all aiming towards a patient-tailored rehabilitation program.

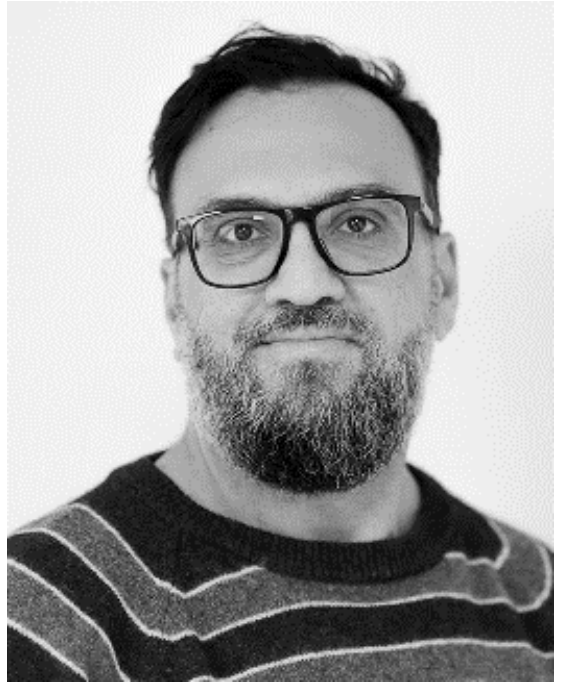

**Ilias Tachtsidis** is a professor at University College London. He leads the Multimodal Spectroscopy and MetaboLight groups in the Biomedical Optics Research Laboratory at the department of Medical Physics and Biomedical Engineering. Tachtsidis is a multidisciplinary scientist with a research portfolio encompassing engineering, physics, computing, neuroscience and clinical medicine. His research focus is technology development in optical neuroimaging (devices, algorithms), through applications (clinical, neuroscience), to data analytics towards generating clinical information and knowledge (computational models).

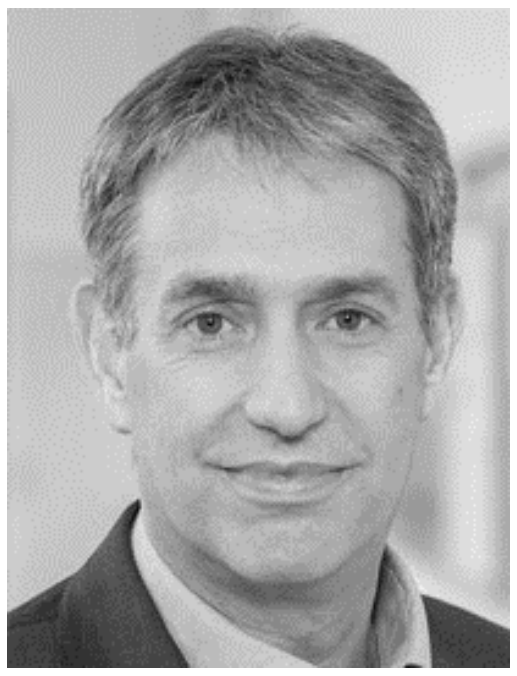

**Daniel Rückert** is Alexander von Humboldt Professor for artificial intelligence (AI) in Medicine and Healthcare at the Technical University of Munich, where he directs the Institute for AI and Informatics in Medicine. He is also a professor in the Department of Computing at Imperial College London. From 2016 to 2020 he served as Head of the Department at Imperial College. His research focuses on medical image computing, data science and AI in medicine.

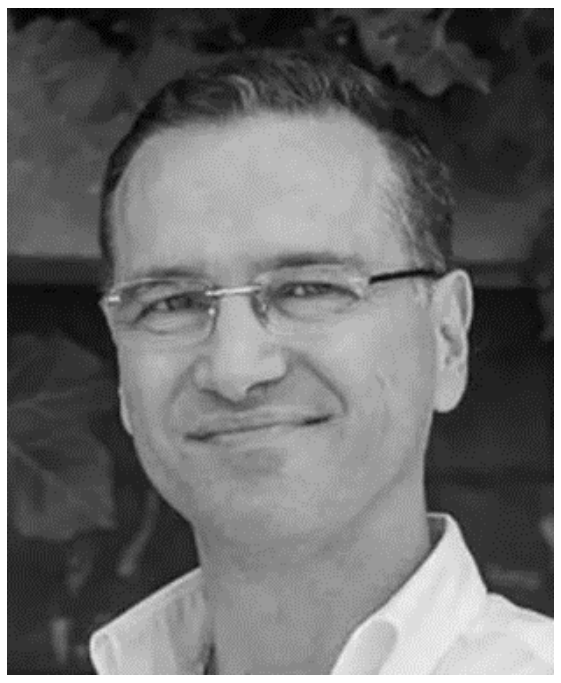

**Francesco Saverio Pavone** is a professor at the University of Florence. He is the director of the Biophotonics area at the European Laboratory for Non-Linear Spectroscopy, focusing on optical microscopy and imaging. Pavone is also coordinator of several EU projects, including the HyperProbe consortium. He participates in the NIH BRAIN initiative and HBP Brain Initiative. He is also the Italian node leader of EBRAIN and the Italian delegate in the Board of Directors of EuroBioImaging.

Biographies and photographs for the other authors are not available.

## Caption List

**Fig. 1** (a) Schematics of HyperProbe1; (b) Picture of the illumination side of HyperProbe1; (c) Picture of the imaging side of HyperProbe1; (d) Testing of the spatial resolution of HyperProbe1; (e) Line profile of the smallest resolved element; (f) Example of glioma biopsy sample.

**Fig. 2** (a) Absorption coefficients for $HbO_2$, HHb, lipids and water, and scattering coefficient of generic brain tissue; b) Molar extinction coefficients of $HbO_2$, HHb, oxCCO and redCCO; c) Simulated fluence rates within the 3D biopsy model at different wavelengths; d) Simulated mean photon pathlength within the 3D biopsy model, as a function of wavelength; e) Partial pathlength distributions of the photons simulated within the 3D biopsy model at various wavelengths.

**Fig. 3** (a) Intercomparison between average reflectance spectra across all samples; (b) Comparison of average reflectance spectra between LGG and HGG samples; (c) Example of spectral frame at 560 nm, acquired with HyperProbe1 on biopsy sample S2; (d) Example of average reflectance spectra in different ROIs of the HyperProbe1 data; (e) Example of average attenuation spectra in different ROIs of the HyperProbe1 data;

**Fig. 4** Spectral fit and corresponding inferred HbT and diffCCO concentration maps of both a HGG (a) and a LGG sample (b), fitting the whole wavelength spectrum (510-900 nm); (c) Histogram showing density distribution of inferred lipid content of each pixel across different LGG and HGG samples; (d) Distribution means of the lipid contents and diffCCO concentrations suggest that lipid mean content could be able to distinguish grading of all samples, whereas no apparent separation is visible for the inferred mean diffCCO concentrations.

**Fig. 5** Spectral fit and corresponding inferred HbT and diffCCO concentration maps of both a HGG (a) and a LGG sample (b), fitting exclusively the NIR range (740-900 nm); (c) Histogram showing density distribution of inferred diffCCO concentrations of each pixel across different LGG and HGG samples; (d) Distribution means of diffCCO and HbO2 concentrations suggest that these parameters could be able to distinguish LGG and HGG samples of glioma tissue, with diffCCO being the most accurate, across all the samples.

**Fig. 6** Histogram showing probability density distribution of inferred oxCCO (a) and redCCO (b) concentrations of each pixel across different LGG (displayed in green) and HGG (displayed in red) grade samples, showing no significance differences in either cases for the range 740-900 nm.

**Table 1** List of components of HyperPRobe1 with specifications.

**Table 2** Technical characteristics and features of HyperProbe1.

**Table 3** Classification of the tissue samples investigated with HyperProbe1.

**Table 4** Composition of the 3D in silico model of brain biopsy used for the MC simulations.